\documentclass[aps,pre,floatfix,twocolumn,showpacs]{revtex4}
\usepackage{latexsym}
\usepackage{graphicx}
\newcommand{\figwidth}{3.275 in}
\usepackage{epsfig}
\usepackage{amsmath}
\usepackage{color}
\begin{document}

\title{Flat histogram diagrammatic Monte Carlo method:
Calculation of the Green's function in imaginary time}
\author{Nikolaos G. Diamantis$^{1}$}
\author{Efstratios Manousakis$^{(1,2)}$}
\affiliation{
$^{(1)}$Department   of    Physics,   University    of   Athens,
  Panepistimioupolis, Zografos, 157 84 Athens, Greece \\
$^{(2)}$ Department  of  Physics and National High Magnetic Field Laboratory,
  Florida  State  University,  Tallahassee,  FL  32306-4350,  USA}

\date{\today}

\begin{abstract}
The diagrammatic Monte Carlo (Diag-MC) method is a numerical
technique which samples the entire diagrammatic series of the
Green's function in quantum many-body systems. In this work, we
incorporate the flat histogram principle in the diagrammatic Monte
method and we term the improved version ``Flat Histogram
Diagrammatic Monte Carlo'' method. We demonstrate the superiority
of the method over the standard Diag-MC in extracting the
long-imaginary-time behavior of the Green's function, without
incorporating any {\it a priori} knowledge about this function, by applying
the technique to the polaron problem.
\end{abstract}
\maketitle

\section{Introduction}
Quantum Monte Carlo methods for bosonic systems continue to
provide very useful insight into the nature of such systems and
can be carried out on very large size systems which allows
extrapolation to the thermodynamic limit. Interacting fermion
systems and frustrated spin systems in more than one dimension,
however, are plagued with the infamous so-called minus-sign
problem, which leads to exponential growth of the statistical
error with the system size. Since most real systems are ``deep
down'' fermionic, including the electronic structure of solids,
and the nuclear, neutron and quark matter, computational physics
faces a bottleneck of paramount importance.

The diagrammatic Monte Carlo (diag-MC)
method\cite{polaronDMC1,polaronDMC2}, which has been
successfully applied to a wide variety of 
problems in a range of fields,
including ultra-cold atoms trapped in optical 
lattices\cite{unitarity}, superconductivity\cite{cooperon},
the Fermi-polaron problem\cite{fermi-polaron},
systems of correlated fermions\cite{DMC-CF,VanHoucke,DMFT-DMC},
in systems of excitons\cite{burovski}, and frustrated quantum 
spin systems\cite{frustrated},
is based on the
interpretation of the entire sum of all Feynman diagrams as an
ensemble averaging process over their corresponding configuration
space. Thus, a Markov sequence $d_0$, $d_1$, ..., $d_n$, ... , can
be defined by means of transitions $d \to d'$ based on the weights
of the particular series of Feynman diagrams, which samples these
diagrammatic terms with the exact relative weights as in the
diagrammatic expansion. One of the great advantages of the method
is that it sums only the linked diagrams and, thus, in the case of
fermion systems, the infamous minus-sign problem could become a
minus-sign ``blessing''\cite{blessing}. 

In addition to the large cancellation due to the interference
terms due to the fermion statistics, which severely limits the
``signal-to-noise'' ratio in MC simulation of fermionic systems,
we also face the problem of extracting the low-energy physics in
most many-body problems. In calculating the Green's function in
imaginary time $\tau$, this problem shows up in our attempt to
estimate the long-time behavior of these functions as a function
of $\tau$. In this limit the Green's function decays rapidly
(because of the other high energy excitations) and, it quickly
becomes very small and the information about the low-lying
excitations gets lost in the noise of numerical data.

The so-called ``flat histogram methods''\cite{berg,oliviera,wang}
have been used to improve the Monte-Carlo simulation of classical
systems, for example for systems undergoing first order phase
transitions, systems with rough energy landscapes, etc. In
addition, the Wang-Landau  algorithm\cite{wang} has been applied to
the simulation of {\it equilibrium} statistical mechanical
properties of {\it quantum} systems\cite{troyer}. More precisely,
Troyer Wessel and Alet (TWA)\cite{troyer} have used the fact that quantum
Monte Carlo simulation uses the mapping of a quantum many-body
system to a classical system which allows them to generalize the
use of the flat histogram method to this case. Using this
generalization, they show that the algorithm becomes efficient by
greatly reducing the tunneling problem in first order phase
transitions and, in addition, the algorithm allows them to
calculate directly the free-energy and entropy. Following this 
idea, Gull {\it et al.}\cite{gull} have applied the idea to the continuous-time 
quantum Monte Carlo approach to the quantum impurity solver needed 
for all dynamical mean-field theory applications. 
While this implementation
can include the use of diagrammatic Monte Carlo\cite{gull},
these calculations proceed by calculating observables 
in an equilibrium-like formulation, i.e., using the ``standard'' density of
states which appears in the partition
function as the distribution which is made flat by the 
application of the WL algorithm.

In this paper, we introduce a
new version of the diag-MC which is superior to standard diag-MC
because it allows us to extract the long imaginary-time (low
energy) behavior of the Green's function $G(\tau)$ accurately
without using any {\it a priori} knowledge about the behavior of
$G(\tau)$. The idea of this new method is based on combining the
principle of flat histograms\cite{berg,wang} and the diag-MC
method. This ``flat histogram diagrammatic Monte Carlo'' (FHDMC)
method allows us to calculate accurately and directly, without
introducing any ``tricks''\cite{polaronDMC2} and with no {\it a
priori} knowledge, the imaginary-time dependence of the Green's
function $G(\tau)$ which varies over many orders of magnitude. The idea is
different from that of TWA and Gull et al.\cite{gull} because, as it will become
clear in the following, TWA make flat the ``density of states''
entering in the partition function, while in our approach the
role of the density of states is selected in such a way to 
make $G(\tau)$ flat. In this
paper, we first introduce the general method of the FHDMC and,
then, the efficiency of the method is demonstrated in the example
of the Fr\"ohlich polaron problem\cite{froehlich,feynman} 
where the standard diag-MC
method has been extensively shown to be
accurate\cite{polaronDMC1,polaronDMC2}.

\section{The method}
First, let us consider a simple case of the diagrammatic Monte
Carlo method in order to illustrate the new idea. The Diag-MC
method is a Markov process which samples an infinite series of the
form:
\begin{eqnarray}
G(\tau) &=& \sum_n^{\infty} I_n(\tau), \hskip 0.2 in I_0(\tau) =  G^0(\tau), \\
I_n (\tau) &= &\int d\vec x_1 d\vec x_2 ... d\vec x_n F_n(\vec
x_1,\vec x_2,...,\vec x_n, \tau),
\end{eqnarray}
where as the order $n$ of the expansion increases, the number of
integration variables increases in a similar manner.

The Diag-MC method is a Markov process $n \to n'$ which generates
the distribution $I_0(\tau)$, $I_1(\tau)$, ..., $I_n(\tau)$, ....
The entire sum $G(\tau)$ can be calculated stochastically if we
know the exact value of one of the terms, say, $I_0(\tau)$. Then,
\begin{eqnarray}
I_n(\tau)={{N_n} \over {N_0}} I_0(\tau),
\end{eqnarray}
where $N_n$ is the number of times the $n^{th}$ term appears in
the Markov sequence. As a result the fluctuations (and the error)
in estimating $I_n(\tau)$ depends crucially on the fluctuations of
$N_0$. As a result if the value of $I_0(\tau)$ relative to other
terms of the series is small, then, the population $N_0$ could be
a small fraction of the total population
$N_T = \sum_{n=0}^{\infty} N_n$,
which is the total number of MC steps, and, thus, the error in the
estimate of $I_n(\tau)$ by using the above expression will be
large. To be more concrete, as the value of $\tau$ increases the
value of $n=n_{max}$ where $I_n$ attains its maximum 
increases, and, $I_{n_{max}}$ increases exponentially with $\tau$. 
 Therefore, for large enough
values of $\tau$ the ratio $I_{n_{max}}/I_0$ becomes many orders
of magnitude larger than unity. This corresponds to the problem of
critical slowing down in classical MC simulation, a problem which
can be addressed by the flat histogram
methods\cite{berg,oliviera,wang}. The flat histogram method
renormalizes these populations by known factors (which can be
easily estimated) and, then, samples a more-or-less flat histogram
of such populations. 

%\subsection{Application of the flat histogram approach}
 We will apply flat histogram
methods\cite{berg,wang} on two different versions of applying the diag-MC.

i) {\it Sampling of $G(\tau)$ for a fixed value of $\tau$:}
This is very similar to the application of flat histogram techniques
applied to classical statistical mechanics. 
 In order to apply any flat histogram technique to our
problem we map the particular value of $n$ to the ``energy'' level
in standard flat histogram methods for classical statistical
mechanics and the integrals $I_n(\tau)$ to the density of states
which corresponds to the corresponding configurations.

ii) {\it Sampling of the histogram of $G(\tau)$:} 
We  divide the range of $\tau$, i.e., $[0,\tau_{max}]$ into $L$
equal intervals $\delta_i = [\tau_i, \tau_{i+1})$, where
$\tau_{i+1}=\tau_i + \Delta \tau$. In this case the variable
$\tau$ is also sampled by the same Markov process.

The histogram $g_l$ of $G(\tau)$ defined as
\begin{eqnarray}
g_l = {{1} \over {\Delta \tau}} \int_{\tau_{l-1}}^{\tau_l} d\tau
G(\tau),
\end{eqnarray}
requires the histograms ${\cal I}^{(l)}_k$ of $I_k(\tau)$ defined
by
\begin{eqnarray}
{\cal I}^{(l)}_k = {{1} \over {\Delta \tau}}
\int_{\tau_{l-1}}^{\tau_l} d\tau I_k(\tau),
\end{eqnarray}
such that
\begin{eqnarray}
g_l = \sum_{k=0}^{\infty} {\cal I}^{(l)}_k
\end{eqnarray}
The FHDMC can be applied in a similar way to the one discussed in
the case for fixed $\tau$. Here, we apply the flat histogram
methods to find the histogram $g_l$ by mapping the value of the
interval $l$ to the ``energy'' level and $g_l$ to the
``density-of-states'' in the classical case.

We have chosen to apply for the first case above both
multicanonical\cite{berg} and the Wang-Landau (W-L)
algorithm\cite{wang}, and for the second case only
the WL algorithm.

The multicanonical algorithm is applied as follows: 
First, for a given fixed value of $\tau$ we carry out an initial 
exploratory run,  where we find that the distribution $I_n$ of the 
values of $n$ peaks
at some value of $n=n_{max}$, which depends on the chosen value of $\tau$.
This distribution falls off rapidly for $n > n_{max}$, and, thus,
we can determine the maximum value $n_{c}$ of $n$ visited by 
the Markov process.
We choose a value of $m$ safely greater than $ n_{c}$, 
such that the value of $I_m$ is practically zero. 
Then, we assign to each one of the $n^{th}$ term  
an initial weight $w^0_n=1$ for all $n=0,1,2...,m$. 
After a certain, relatively small
 number $N_0$ of diag-MC steps,  we redefine the weights to 
$w^1_n = w^0_n/H^0_n$ for all $n=0,1,2...,m$, where
$H^0_n$ is the height of the distribution of configurations
 of the $n^{th}$ order. In the next step, we carry out  $N_0$
 diag-MC steps after which we change the weights to
 $w^2_n=w^1_n/H^1_n$ for all $n$ less or equal to $m$, 
where $H^1_n$ is the height of the new distribution of
configurations of the $n^{th}$ order. This procedure is repeated several 
times, thus, defining a sequence of histograms $H^k_n$ and of
weights $w^k_n$;
at the end of each one of these steps k, we determine whether or not
the histogram $H^k_n$ of $n$ is ``flat'', within a some acceptable tolerance
level.  When, the histogram $H^k_n$ becomes ``flat'' at the $k=\bar k$ step
we stop this process, and we begin a Markov process for a 
relatively large number $N$ of MC steps, by re-weighting the acceptance
rates using the weights $w^{\bar k}_n$ determined during the last
$k=\bar k$ process, and this way we determine $G(\tau)$.

The W-L algorithm is applied as follows:

Using the standard DMC algorithm (to be discussed for the
case of our version of the polaron problem in Sec.~\ref{DMCA})
we sample the modified density of states $F_n/\rho(n)$, where
$n=0,1,2,...,m$ and $F_n$ is $I_n$ when we sample $I_n$ (case i) above), 
and $F_n=g_n$,  $n=1,...,L$,  when we sample the histogram of 
$G(\tau)$ (case ii) above). 
Initially, we take $\rho(n)=1$ for all values of $n$. 
Every time a state $k$ appears in the
Markov process the value of $\rho(k)$ is modified immediately
in the next step to $\rho(k) \to \rho(k) f$ where the factor $f>1$.
After a certain number of iteration $N_0$ we check whether or not
the histogram of the states is flat within a tolerance of our
specification.  If the histogram is not flat, we repeat the previous
step for another $N_0$ iterations. If the histogram is found to be
flat, then, we set the values of the histogram to zero and we repeat the
whole previous process by using $f \to \sqrt{f}$.
We stop the process when $f<f_{min}$, where $f_{min}$ is 
our pre-specified value. The final values of $\rho(n)$ after normalization
are the density of states, which are proportional to the
values of $F_n$.

\section{Application to the polaron problem}

The standard Diag-MC has been extensively applied to the
Fr\"ohlich polaron Hamiltonian which describes a single electron in a phonon
field:
\begin{eqnarray}
H &=& \sum_{{\bf k}} e(k) a^{\dagger}_{{\bf k}} a_{{\bf k}} +
\omega_0 \sum_{\bf k} (b^{\dagger}_{{\bf k}} b_{{\bf k}} + {1
\over 2})
\nonumber \\
 &+& \sum_{\bf k,q} V(q)(b^{\dagger}_{\bf q} - b_{-{\bf
q}})a^{\dagger}_{{\bf k-q}} a_{{\bf k}},\\
 V({\bf q}) &=& {i \over {\sqrt{\Lambda}}}  \sqrt{2 \sqrt{2}
\alpha \pi} {1 \over q}, \hskip 0.2 in e(k) = {{k^2} \over 2}
\end{eqnarray}
where the summation extends over the first Brillouin zone and
$\Lambda$ is the volume of the system. This Hamiltonian takes into
account a single optical phonon with fixed frequency $\omega_0$
and $a^{\dagger}_{\bf k}$ and $b^{\dagger}_{\bf k}$ are electron
and phonon creation operators. Here, we will work in imaginary
time at $T=0$. The free electron and phonon propagators are
respectively given by
\begin{eqnarray}
G^0({\bf k}, \tau_2-\tau_1) &=& e^{-e(k) (\tau_2 - \tau_1)}
\Theta(\tau_2 - \tau_1),\\
D(q,\tau_2-\tau_1) &=& e^{-\omega_0 (\tau_2-\tau_1)}.
\end{eqnarray}

%\subsection{Soluble case}

\begin{figure}
\vskip 0.5 in \epsfig{file=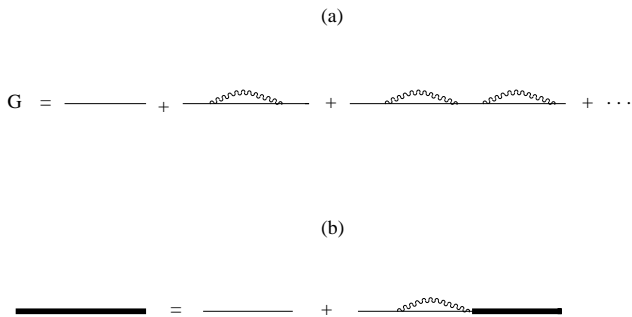,width=\figwidth}
\caption{(Color online) (a) An infinite series of selected
diagrams contributing to the single electron Green's function.
(b) The Dyson equation which sums the selected series.
The wiggly line denotes a phonon propagator, while a thin (thick) solid line
denotes the bare (renormalized) fermion propagator.}
\label{fig:0} \vskip 0.5 in
\end{figure}

While the Diag-MC algorithm and our computer program are quite
general and can solve exactly (within statistical errors) the
single polaron problem, we choose to restrict ourselves to
sampling only the diagrams of the infinite series shown in
Fig.~\ref{fig:0}. The reason is that we can sum up this infinite
series of selected diagrams exactly and this allows us to compare
the results of the standard Diag-MC and the FHDMC with the exact
solution. This series is summed by solving Dyson's equation which
for the case of $k=0$ is given as:
\begin{eqnarray}
G(\tau) &=& G^0(\tau) + \int_0^{\tau} d\tau_2 \int_0^{\tau_2}
d\tau_1 G^0(\tau_1) \Sigma(\tau_2-\tau_1) \nonumber \\
&\times& G(\tau-\tau_2), \\
\Sigma(\tau') &=& 2\alpha \sqrt{2} \pi \int {{d^3q} \over {(2\pi)^3}} {1
\over {q^2}} e^{-(q^2/2+1)\tau'},
\end{eqnarray}
where we have taken $\omega_0=1$. This equation can
be transformed in Laplace's space, using the properties of
convolution integrals, as follows:
\begin{eqnarray}
{\tilde G}(s) = {\tilde G}^0(s) + {\tilde G}^0(s) {\tilde
\Sigma}(s) {\tilde G}(s), \\
{\tilde G}^0(s) = {1 \over s}, \hskip 0.2 in 
{\tilde \Sigma}(s) = {\alpha \over {\sqrt{s+1}}}, \hskip 0.2 in \Re(s)
> -1
\end{eqnarray}
which yields
\begin{eqnarray}
{\tilde G}(s)= {{\sqrt{s+1} \over {s\sqrt{s+1}-\alpha}}}.
\end{eqnarray}
The inverse Laplace transform is given by
\begin{eqnarray}
G(\tau) = {1 \over {2\pi i}} \lim_{T \to \infty} \int_{\sigma-i
T}^{\sigma+i T} {\tilde G}(s) e^{s \tau} ds, \hskip 0.3 in \sigma
> 0.
\end{eqnarray}
Using the corresponding Bromwich contour we obtain:
\begin{eqnarray}
G(\tau) &=& A_0 e^{s_1 \tau} + {{e^{-\tau}} \over {2\pi}}
\int_0^{\infty} dr {{2\alpha \sqrt{r} e^{-r\tau}} \over {r(r+1)^2 + \alpha^2}},\\
A_0 &=& {{s_1^2 + s_1 + \alpha \sqrt{s_1 + 1}} \over {(s_1 - s_2)(s_1-s_3)}},
\end{eqnarray}
where  $s_1, s_2, s_3$ are the three roots of the equation $s^3 + s^2 - 
\alpha^2 =0 $, $s_1$ is the real root and $s_2$ and $s_3$ are complex conjugate
to each other. For $\alpha = 2$ we find that
$s_1=1.314596$  and $A_0=0.7788386$. Asymptotically for $\tau \to
\infty$ we have
\begin{eqnarray}
G(\tau) = A_0 e^{s_1 \tau}.
\end{eqnarray}
Therefore, the polaron ground state energy in the approximation
given by the series in Fig.~\ref{fig:0} is $E_0 = -s_1$. For a
finite chemical potential $\mu$, $G(\tau)$ should be modified to
$A_0\exp((s_1+\mu)\tau)$.

\section{Diagrammatic Monte Carlo algorithm}
\label{DMCA}

{\it Fixed $\tau$ diag-MC}: 

As discussed earlier we have applied
the diag-MC for fixed $\tau$. The fixed-$\tau$ diag-MC is very
similar to the approach outlined by Prokof'ev {\it et
al.}\cite{polaronDMC1}. It includes only transitions from order $n$
to $n+1$ where we select the beginning $\tau_1$, the end $\tau_2$
and the momentum ${\bf q}$ of the phonon propagator. In addition, it
includes transitions from $n$ to $n-1$ (except when $n=0$; this
exception modifies the acceptance rate from $0 \to 1$ compared to
all other transitions originating from $n\ne 0$) where the phonon
propagator is removed. We modify this part of the general
simulation algorithm in order to simulate only the series in
Fig.~\ref{fig:0} as follows: when we attempt the MC move $n \to
n+1$, i.e., we attempt to add a new phonon propagator at time
instants $\tau^{(n+1)}_1$ and $\tau^{(n+1)}_2$, the free fermion
line where the instant $\tau^{(n+1)}_1$ is to be inserted is
selected to be one of those existing free-fermion lines which do
not have the same ends with one and the same phonon propagator,
and $\tau^{(n+1)}_2$ is selected to be between $\tau^{(n+1)}_1$
and the end of the selected free-fermion line.

{\it Sampling of the histogram of $G(\tau)$:} 

As discussed earlier
in this case we consider the histogram in the intervals
$\delta_i$. Using the behavior of $G(\tau)$ for large $\tau$ we
calculate the ground state energy. The updating procedure between
different orders $n$ is the same as in the case of fixed $\tau$ discussed
in the previous paragraph.
In addition, here, we allow transitions to a different time
$\tau'$ which may belong to a different interval. The value of
$\tau'$ is selected as in Ref.~\onlinecite{polaronDMC1}, with the
only difference that $\tau'$ is selected in the interval
$[\tau_{end},\tau_{max}]$, instead of the interval
$[\tau_{end},\infty]$ (where $\tau_{end}$ is the end of the last
phonon propagator in the particular diagram).

\section{Results and comparison}

\begin{figure}
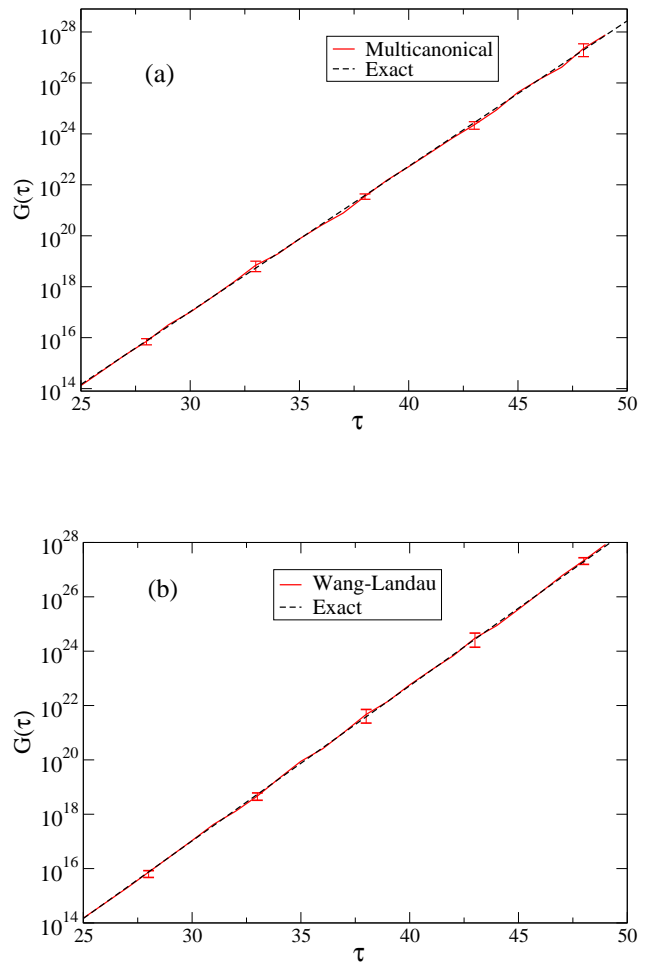

\vskip 0.5 in 
\epsfig{file=Fig2a.eps,width=\figwidth} \\
\vskip 0.5 in \epsfig{file=Fig2b.eps,width=\figwidth}
\caption{(Color online) Results of FHDMC for the case of fixed
$\tau$ to the reduced polaron problem. 
Top: The results of FHDMC obtained with the application of
multicanonical\cite{berg}. Bottom: The results of FHDMC obtained with the application of Wang-Landau\cite{wang} algorithm for the same number of 
iterations (which approximately leads to the same
CPU time).  In addition, the exact
results are shown for comparison.}
\label{fig:1}
\vskip 0.5 in
\end{figure}
Fig.~\ref{fig:1} presents the results of our calculation of
$G(\tau)$ for fixed values of $\tau$. The results of FHDMC were
obtained with the application of multicanonical\cite{berg} (top of Fig.~\ref{fig:1}) and
Wang-Landau\cite{wang} (W-L) algorithm (bottom of Fig.~\ref{fig:1})
for the reduced polaron problem.
We calculated $G(\tau)$ for $\tau=25.0,26.0,...,49.0$ and for approximately the
the same amount of CPU time for comparison. 

In our multicanonical simulation 
for a given value of $\tau$, our tolerance for the value of histogram of
$I_n(\tau)$ for any $n$ was lower than twice and higher than half the
value of the histogram for $I_0(\tau)$. Renormalization of the ``density 
of states''  was performed every $10^5$ iterations. After 
the histogram became flat within our tolerance, we carried out
$3 \times 10^7$ iterations for each instant of time considered.  
The total number of iterations required was $8 \times 10^8$.

In our simulation using the W-L algorithm, we gave the modification 
factor $f$ the initial value $f=e$ and we reduce it according to 
$f_{i+1} = \sqrt{f_i}$ every time the histogram became ``flat'' within 
our tolerance.  Our tolerance for ``flatness'' was such that
the absolute value of the difference of the histogram $H_n$ 
for each value of $n$ from  the average value of the histogram  
$\bar H$ to be less than $\bar H/10$.
The ``flatness'' of the histogram was assessed every $10^5$ iterations.
Choosing the final value for $f$ to be  1.000007629,  the 
total number of iterations required was $8 \times 10^8$.

Notice that both flat histogram algorithms yield
$G(\tau)$ for any $\tau$ including very long-$\tau$
where $G(\tau)$ varies over many orders of magnitude. By fitting the long
imaginary-time part of the Green's function to a single exponential
$G(\tau)= Z e^{-E_{0} \tau}$, we find that in the case of
multicanonical the ground state energy is $E_{0} = -1.310\pm 0.004$
and $Z=0.88\pm 0.11$, while in the case of the W-L algorithm
$E_{0} = -1.311 \pm 0.004$ and $Z=0.88 \pm 0.12 $. The exact
values are $E_{0}=-1.3146$ and $Z=0.77883$.

\begin{figure}
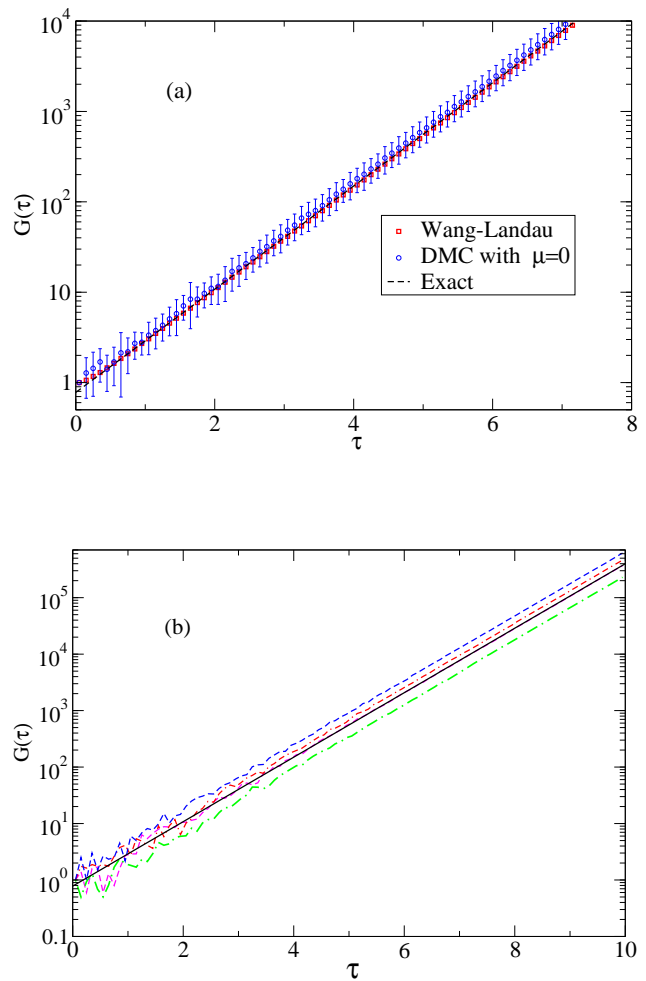

\vskip 0.5 in \epsfig{file=Fig3a.eps,width=\figwidth} \\
\vskip 0.5 in \epsfig{file=Fig3b.eps,width=\figwidth} 
\caption{(Color online) (a) Comparison of the histogram of $G(\tau)$
obtained using the standard diag-MC, and the FHDMC. In addition,
the exact $G(\tau)$ is shown for comparison. (b) The results of 
four different MC runs are compared with the exact results (solid line). 
See text for details.} \label{fig:2} \vskip
0.5 in
\end{figure}

\begin{figure}
\vskip 0.5 in \epsfig{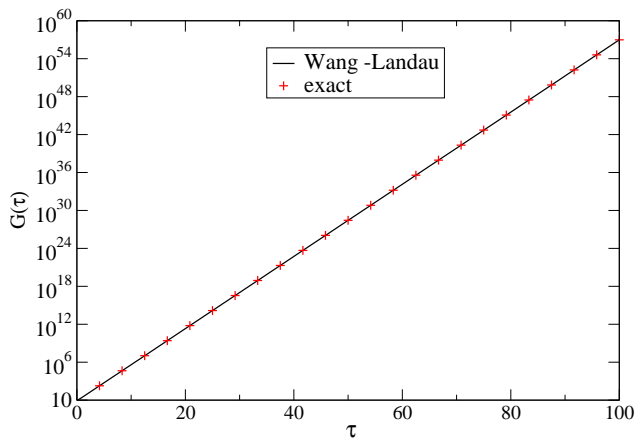}
\caption{(Color online) Comparison of the exact histogram of $G(\tau)$
with that obtained using  FHDMC using the WL algorithm to make 
the histogram flat. The error bars are too small to be seen on the present
scale and have been omitted for clarity.
Notice that we have used 2000 time intervals for the histogram. } \label{fig:3} \vskip
0.5 in
\end{figure}

Fig.~\ref{fig:2}(a)  compares the results
of our FHDMC for $g_l$ obtained with the W-L algorithm with the
results of the standard implementation of the diag-MC without using
the so-called guidance exponential function (i.e., 
parameter $\mu=0$). 
The simulation using the W-L algorithm was done as described earlier
when discussing the results of Fig.~\ref{fig:1} for the
calculation of $G(\tau)$ for constant $\tau$. 
For comparison the diag-MC method was also applied for approximately
the same CPU time. Notice that while the WL simulation gives results with
negligible error bar (its size in this Figure is about the size of
the symbols), the diag-MC calculation has significant errors. 
However, if we look closely we notice that the average values, from
each time window to the next, do not fluctuate in a way which is
consistent with the size of the error bar. In order to clarify this,
we provide Fig~\ref{fig:2}(b), where the results of four different runs 
are presented (shown with different types of lines) where each curve is obtained
using the same number of iterations starting from 
a different random initial configuration. Notice that, from one run
to the other, the results
for the histogram of $G(\tau)$ move in a correlated way (i.e., together)  
up or  down and the four curves are parallel to 
each other for large values of $\tau$. 
The reason for that is the following.  The exponential 
growth of $G(\tau)$ with $\tau$ 
makes $g_n >> g_1$ for large $n$. As a consequence of that 
the number  of Monte Carlo steps ``falling'' in the
first interval is relatively small. 
This means that the statistical fluctuations of the histogram height of 
the first interval are much larger than
the statistical fluctuations of the histogram height of the other intervals. 
Within the diag-MC method the value of $g_n$ is obtained by simply 
multiplying the ratio of the corresponding histogram heights $H_n/H_1$ with the 
value of $g_1$, which is analytically known. 
Thus, due to these fluctuations,  the value of $H_1$ is significantly
different between two different MC runs and this affects the entire
histogram by a multiplicative factor.

The problem discussed in the previous paragraph has been addressed 
by the standard diag-MC method when applied to the 
polaron problem by using an exponential guidance function\cite{polaronDMC1}. 
The diag-DMC, as has been applied in the polaron problem\cite{polaronDMC1},
uses the chemical potential as a ``tunable parameter'' to reduce the above
mentioned statistical variance.   In this particular problem,
 the exact asymptotic 
solution for $G(\tau)$ is a single exponential, $G(\tau) = Z
\exp(-E_0 \tau)$, where $E_0$ is the ground state energy.
Therefore, 
using the chemical potential $\mu$, which corresponds to multiplying 
$G(\tau)$ by an exponential
factor $\exp(\mu \tau)$ with $\mu$ as a tunable parameter, is effectively 
 a simple way to make the histogram flat at long-$\tau$. The histogram
becomes flat for the polaron problem simply when we choose $\mu \sim E_0$. 
If we use a value of $\mu$ close to
$E_0$ this will improve the statistics of the diag-MC significantly.
The closer the value of $\mu$ is tuned to the exact value of $E_0$, the
better the diag-MC works.
Therefore, in view of the present work, this
works because such a guidance function makes the histogram of
$G(\tau)$ approximately flat. 
However, this method of using
such a guidance function is inferior to the present method because
of the following reasons.

We do not {\it a priori} know the guidance function for more
complex problems. An exponential decay of $G(\tau)$ with a single
characteristic exponent is not the general rule in interacting
many-body systems. In such systems, the general rule is that
the Green's function in imaginary time can be expressed as
\begin{eqnarray}
G(\tau) = \int_0^{\infty} d\omega A(\omega) e^{-\omega \tau}
\end{eqnarray}
where $A(\omega)$ is related to the analytic continuation of the
spectral function in imaginary time. Therefore, in general, we
expect to need a continuum of such exponential
energy scales. In most interesting systems it is not simple
to extract $A(\omega)$ by the approach of tuning different
exponential factors used in Ref.~\cite{polaronDMC1,polaronDMC2}.
The method presented here,
however, constructs the exact ``density of states'', with 
no {\it a priori} knowledge about it.

In order to further demonstrate the power of the present method,
in Fig.~\ref{fig:3} we present the calculated histogram of $G(\tau)$
in a detailed mesh of 2,000 intervals ranging from 0 to 100. This calculation
took approximately 3 days of CPU time on a single 3 GHz processor.
Notice that the agreement with the exact solution is excellent over
approximately 60 orders of magnitude! 

%\begin{table}
%\begin{center}
%\begin{ruledtabular}
%\caption{}
%\begin{tabular}{c c c c c c }
%\label{tab:results}
%   &  SDMC  & FHDMC-WL & Exact          \\
%\hline  $E_0$ & $-1.3211 \pm 0.0005$ & $-1.3167 \pm 0.0004$ &
%-1.314596
%\\
%Z & $0.78 \pm 0.02$ & $0.73 \pm 0.01$ & 0.77883

%\end{tabular}
%\end{ruledtabular}
%\end{center}
%\end{table}

\section{Conclusions}

In conclusion,  we have
presented the FHDMC method to improve the standard diag-MC based
on the idea of flat histogram methods. This idea was demonstrated on
an exactly soluble problem, however, our method is very general
and can be implemented for any problem where the diagrammatic MC
method can be applied. We have shown with this method that the
statistical fluctuations can be controlled even for very large
values of the imaginary time without the need for any {\it a priori}
knowledge about the behavior of $G(\tau)$.

%\section{Acknowledgments}

\end{document}